
%
%
%
%
\magnification=1200\overfullrule=0pt\baselineskip=15pt
\vsize=22truecm \hsize=15truecm \overfullrule=0pt\pageno=0
\font\titlefont=cmbx12 scaled \magstep1
\font\sectnfont=cmbx12 scaled \magstep0
\def\mname{\ifcase\month\or January \or February \or March \or April
           \or May \or June \or July \or August \or September
           \or October \or November \or December \fi}
\def\date{\hbox{\strut\mname \number\year}}
\def\scnum{\hbox{SPhT-93/022\strut}}
\def\banner{\hfill\hbox{\vbox{\offinterlineskip
                              \scnum\date}}\relax}
\footline={\ifnum\pageno=0{}\else\hfil\number\pageno\hfil\fi}
%
%
%
\newcount\FIGURENUMBER\FIGURENUMBER=0
\def\fig#1{\expandafter\ifx\csname FG#1\endcsname\relax
               \global\advance\FIGURENUMBER by 1
               \expandafter\xdef\csname FG#1\endcsname
                      {\the\FIGURENUMBER}\fi
           Fig.~\csname FG#1\endcsname\relax}
\newcount\REFERENCENUMBER\REFERENCENUMBER=0
\def\reftag#1{\expandafter\ifx\csname RF#1\endcsname\relax
               \global\advance\REFERENCENUMBER by 1
               \expandafter\xdef\csname RF#1\endcsname
                      {\the\REFERENCENUMBER}\fi
             \csname RF#1\endcsname\relax}
\def\ref#1{\expandafter\ifx\csname RF#1\endcsname\relax
               \global\advance\REFERENCENUMBER by 1
               \expandafter\xdef\csname RF#1\endcsname
                      {\the\REFERENCENUMBER}\fi
             [\csname RF#1\endcsname]\relax}
\def\refto#1#2{\expandafter\ifx\csname RF#1\endcsname\relax
               \global\advance\REFERENCENUMBER by 1
               \expandafter\xdef\csname RF#1\endcsname
                      {\the\REFERENCENUMBER}\fi
           \expandafter\ifx\csname RF#2\endcsname\relax
               \global\advance\REFERENCENUMBER by 1
               \expandafter\xdef\csname RF#2\endcsname
                      {\the\REFERENCENUMBER}\fi
             [\csname RF#1\endcsname--\csname RF#2\endcsname]\relax}
\def\refand#1#2{\expandafter\ifx\csname RF#1\endcsname\relax
               \global\advance\REFERENCENUMBER by 1
               \expandafter\xdef\csname RF#1\endcsname
                      {\the\REFERENCENUMBER}\fi
           \expandafter\ifx\csname RF#2\endcsname\relax
               \global\advance\REFERENCENUMBER by 1
               \expandafter\xdef\csname RF#2\endcsname
                      {\the\REFERENCENUMBER}\fi
            [\csname RF#1\endcsname,\csname RF#2\endcsname]\relax}
\newcount\EQUATIONNUMBER\EQUATIONNUMBER=0
\def\EQNO#1{\expandafter\ifx\csname EQ#1\endcsname\relax
               \global\advance\EQUATIONNUMBER by 1
               \expandafter\xdef\csname EQ#1\endcsname
                      {\the\EQUATIONNUMBER}\fi
            \eqno(\csname EQ#1\endcsname)
                      \relax}
\def\eq#1{\expandafter\ifx\csname EQ#1\endcsname\relax
               \global\advance\EQUATIONNUMBER by 1
               \expandafter\xdef\csname EQ#1\endcsname
                      {\the\EQUATIONNUMBER}\fi
          Eq.~(\csname EQ#1\endcsname)\relax}
\def\eqand#1#2{\expandafter\ifx\csname EQ#1\endcsname\relax
               \global\advance\EQUATIONNUMBER by 1
               \expandafter\xdef\csname EQ#1\endcsname
                        {\the\EQUATIONNUMBER}\fi
          \expandafter\ifx\csname EQ#2\endcsname\relax
               \global\advance\EQUATIONNUMBER by 1
               \expandafter\xdef\csname EQ#2\endcsname
                      {\the\EQUATIONNUMBER}\fi
         Eqs.~\csname EQ#1\endcsname{} and \csname EQ#2\endcsname\relax}
\def\eqto#1#2{\expandafter\ifx\csname EQ#1\endcsname\relax
               \global\advance\EQUATIONNUMBER by 1
               \expandafter\xdef\csname EQ#1\endcsname
                      {\the\EQUATIONNUMBER}\fi
          \expandafter\ifx\csname EQ#2\endcsname\relax
               \global\advance\EQUATIONNUMBER by 1
               \expandafter\xdef\csname EQ#2\endcsname
                      {\the\EQUATIONNUMBER}\fi
          Eqs.~\csname EQ#1\endcsname--\csname EQ#2\endcsname\relax}
\def\avg#1{\ifmmode\langle#1\rangle\else$\langle#1\rangle$\fi}
%
\begingroup\titlefont\obeylines
\hfil The free energy of the Potts model :\hfil
\hfil from the continuous \hfil
\hfil to the first-order transition region.\hfil
\endgroup\bigskip

\medskip
\centerline{ T. Bhattacharya\footnote{*}%
{Present address: MS B285, Group T-8, Los Alamos National Laboratory,
NM 87544, U.S.A.}, R. Lacaze\footnote{**}%
{Chercheur au CNRS} and A. Morel}
\centerline{Service de Physique Th\'eorique de Saclay\footnote{***}%
{Laboratoire de la Direction des Sciences de la Mati\`ere du CEA }}
\centerline{91191 Gif-sur-Yvette Cedex, France}

\bigskip\bigskip\bigskip\centerline{{\sectnfont ABSTRACT}}\medskip
We present a large $q$ expansion of the 2d $q$-states Potts model free
energies up
to order 9 in $1/\sqrt{q}$. Its analysis leads us to an ansatz which,
in the first-order region, incorporates properties inferred from the known
critical regime at $q=4$, and predicts, for $q>4$, the $n^{\rm th}$ energy
 cumulant scales as the power $(3 n /2-2)$ of the correlation length.
The parameter-free energy distributions reproduce accurately, without
reference to any interface effect,  the numerical data obtained
in a simulation for $q=10$ with lattices of linear dimensions up to $L=50$.
The pure phase specific heats are predicted to be much larger, at $q\leq10$,
than the values extracted from current finite size scaling analysis of extrema.
Implications for safe numerical determinations of interface tensions are
discussed.

\bigskip\bigskip\bigskip\bigskip\banner
\bigskip\noindent Submitted for publication to {\sl Europhysics Letters}
\vfil\eject

Much effort has been recently devoted to the 2-d Potts model, both with
numerical and analytical techniques. In the former case, the goal was
either to test numerical
algorithms and criteria for distinguishing first-order from continuous
transitions, or to learn how to extract previously unknown quantities such as
the interface tension.
Although much progress was accomplished, there remains some unsatisfactory
issues such as, for example, slight inconsistencies in finite size scaling
analysis of the energy cumulants close to the transition temperature
$\beta_{\rm t}^{-1}$, and discrepancies between exact results and numerical
simulations for the interface tension.
This question is important since only numerical simulations can determine this
quantity in other cases of physical interest such as the 3-d $q=3$ Potts model
or QCD at the deconfinement transition.

Analytical works have shown \ref{boko} that close to $\beta_{\rm t}$, the
partition function $Z$ of the Potts model, in a box of volume $V=L^2$ with
periodic boundary conditions (used all through in this work),
is equal to the sum of the `partition functions' $Z_i$ of the $q+1$
pure phases, and a term that falls of exponentially faster in the linear
size of the system. We shall presently ignore
this latter term which contains, amongst others, the interface tension effects,
and concentrate on the $i^{\rm th}$ phase free energy
$F_i= \ln Z_i/V$, which is $V$ independent and differentiable  many times with
 respect to the inverse temperature $\beta$ at $\beta_{\rm t}$.

In this letter, we construct explicit formulae for the ordered and disordered
free energies $F_{\rm o}$ and $F_{\rm d}$ of the 2-d $q$-states Potts model
at $q>4$. At large $q$ we perform their expansion in power of $1/\sqrt{q}$.
At low $(q-4)$ we conjecture their behaviour from their known critical
behaviour at $q=4$. We show that these two descriptions match in a large
intermediate $q$ region. Then, using the above rigorous result, we add up the
$Z_i$'s so obtained and make absolute predictions on energy probability
densities in very good agreement with numerical data.
We show that the difficulties encountered in the finite size
scaling analysis of numerical data are due to very large high order cumulants,
which has consequences for the extraction of interface tensions.

Many properties of the model are known exactly \ref{wu}.
In particular, it exhibits a temperature driven phase
transition which occurs at $\beta_{\rm t}=\ln(\sqrt q+1)$.
The transition is second-order for $q \leq 4$, and its critical properties
are described, e.g., by the $\alpha$ and $\nu$ indices, which at $q=4$
take the common value 2/3. Accordingly, the correlation length and the
specific heat there diverge as
$$\xi_{q=4}\sim \mid \beta -\beta_{\rm t} \mid ^{-2/3}\qquad,\qquad
C_{q=4}\sim \mid \beta -\beta_{\rm t} \mid ^{-2/3}. \EQNO{xi4}$$
Hence in the vicinity of $\beta =\beta_{\rm t}$, the ratio ${C / \xi}$ remains
finite at $q=4_-$.

The first-order transition region is $q>4$. There the energies
$E_{\rm o}$ and $E_{\rm d}$ of the ordered and disordered phases
respectively are exactly known at $\beta_{\rm t}$ \ref{baxt}. Recent
works \refand{xi}{bogs} on the largest correlation
length at $\beta_{\rm t}$ have shown a common behaviour as $q\rightarrow 4_+$
$$\xi={1\over 8\sqrt{2}} \ x \ (1+{\cal O}(x^{-2}))\qquad {\rm with} \qquad
x=\exp({\pi^2\over 2 \ln{1\over 2}(\sqrt{q}+\sqrt{q-4})}). \EQNO{xi}$$
These formulae show not only that $\xi$ rapidly diverges as
$q\rightarrow 4_+$, but also that the leading behaviour of \eq{xi} {\it is
accurate over a very wide range of $q$ values}. For example the correction
term in \eq{xi} is still of the order of 1\% for $q$ as large as 75.
The pure phase specific heats $C_{\rm o}$ and $C_{\rm d}$ are unknown,
but their known difference vanishes when $q\rightarrow 4_+$ as
$x^{-{1\over2}}$. These properties are of course in accordance with
the point ( $q=4, \beta=\beta_{\rm t}$ )
being a second-order transition point, and lead one to speculate that
$C_{\rm d}\sim C_{\rm o}$ diverges as $q\rightarrow 4_+$, possibly
in such a way that the ratio $C / \xi$ is finite on both sides of $q=4$.

First we shall take advantage of the fact that, for the correlation length
and latent heat, the "small $q-4$ " region extends in practice up to large
$q$ values, and start from the opposite end. We compute the free energy of
 the model in the framework of a large $q$ expansion, extrapolate down in
$q$ as far as we can, and analyze the resulting energy cumulants as functions
of $x$ in an intermediate $q$-value region. Nice regularities emerge, among
which a smooth behaviour of $C / x$ is ascertained.

The large $q$ expansion of the ordered free energy $F_{\rm o}$ (that for
$F_{\rm d}$ in the disordered phase follows from duality), was obtained
through the Fortuin-Kasteleyn \ref{kast} representation of the Potts model
partition function
$$Z=\sum_{X} ({\rm e}^{\beta}-1)^l q^n \EQNO{FK}$$
where $X$ is any configuration of bonds on a cubic
lattice, $l$ its number of bonds and $n$ its number of connected components,
or clusters of sites ( two sites bound to each other belong to the same
cluster, an isolated site is a cluster). The completely ordered
configuration corresponds to $n=1$ and $l=2 V$.
So the partition function can be reorganized as an
expansion in $q^{-{1\over2}}$ about this configuration:
$$Z_{\rm o} = q ({\rm e}^{\beta}-1)^{2V} \sum_{l\geq 0,n\geq 0} N_{l,n}(V)
    ({{\rm e}^{\beta}-1\over\sqrt{q}})^{-l}q^{n-{l\over2}}, \EQNO{Zo}$$
where $N_{l,n}(V)$ is the number of configurations in a volume $V$ with $l$
removed bonds and $n+1$ clusters.
The enumeration of all the $N_{l,n}(V)$  such that $(l-2n)\leq M$
yields an expansion of $Z_{\rm o}$ to order $M$.
Details will be given elsewhere \ref{futur}.
To any given finite order $M$, a large enough volume $V$ can be chosen to
eliminate all  boundary terms, so that all
configurations retained correspond to {\it disordered islands} in a bulk
{\it ordered phase}. We check that the sum in \eq{Zo} exponentiates in $V$
up to terms of order $M+1$, defining a series for $F_{\rm o}$
truncated beyond order $M$. We have computed up to order $M=9$,
including terms up to $N_{49,20}(V)$.
This series, whose first terms can be compared to existing low
temperature series \ref{enting}, provides us with similar series for the
$k^{\rm th}$ derivative with respect to $\beta$, $F_{\rm o}^{(k)}$.
At $\beta=\beta_{\rm t}$  the $k=0$ (free energy) and $k=1$ (internal energy)
series match the exact results \ref{baxt} up to $M=9$.
The $k=2 \ {\rm and} \ 3$ cases give
$$F_{\rm o}^{(2)}={16\over q}+{34\over q^{3/2}}+{114\over q^2}
+{254\over q^{5/2}} +{882\over q^3}+{1944\over q^{7/2}}
+{6128\over q^4}+{13550\over q^{9/2}} , \EQNO{Foq}$$
$$F_{\rm o}^{(3)}=-{64\over q}-{430\over q^{3/2}}-{2654\over q^2}
-{12186\over q^{5/2}}-{57018\over q^3}-{224732\over q^{7/2}}
-{888024\over q^4}-{3164682\over q^{9/2}} .  \EQNO{Fdq}$$
Let us make a few comments on these expansions.
\item{(i)} $F_{\rm o}^{(2)}$ gives the specific heat
 $C_{\rm o}=\beta_{\rm t}^2 F_{\rm o}^{(2)}$. The $F_{\rm o}^{(3)}$ series
gives -- \avg{(E-E_{\rm o})^3}, the first odd moment of the
energy distribution in this phase.
\item{(ii)} All terms in each series have the same sign, $F^{(2)}$
being of course positive while $F^{(3)}<0$ means that $E>E_{\rm o}$ is
 favoured with respect to $E<E_{\rm o}$.
\item{(iii)} The coefficients are fastly increasing with the order,
the more so for larger values of $k$. This confirms the expectation that
huge energy fluctuations are associated with the large correlation
length, of order $x$ in \eq{xi} when $q$ decreases towards $q=4$.
This is expected to have a direct impact on the numerical analysis of these
models, large values of the high cumulants invalidating the commonly used two
gaussian formula \ref{challa}.
\bigskip
To proceed with a quantitative analysis of $F^{(2)}$ and $F^{(3)}$ as
functions of $q$, we conjecture for these quantities an essential
singularity at $q=4$ as $\xi$ has and construct Pad\'e approximants for the
series of $\ln F^{(k)}$ instead of $F^{(k)}$, i.e. take as estimates
of $F^{(k)}$
$$F_{\rm est}^{(k)}=\exp \bigl[ \hbox{ Pad\'e}(\ln F^{(k)}) \bigr ]
   \EQNO{FPad}$$
The result of this construction for $F^{(2)}$ and $F^{(3)}/F^{(2)}$
as functions of $x$ for $q=30,20,15,10,8,7 \ {\rm and} \ 6$ is summarized in
\fig{1} as a log-log plot.
The error bars are rough estimates of the uncertainties resulting from the
higher order terms and have been obtained by varying the degrees of the
numerator and denominator of the Pad\'e approximant. The lines represent
 our prejudices $F^{(2)}\simeq {\rm Cst} \ x$ and
$F^{(3)}/F^{(2)}\simeq {\rm Cst} \ x^{3/2}$ (see below), where the constants
are fixed by the $q=10$ values.
It is clear that the general trend of both quantities is well reproduced
over an astonishingly large range by such simple forms.

As a by-product of this study we obtain analytical estimates for the ordered
phase specific heat, which are compared to existing numerical data in Table 1.
\vskip 13.5 truecm
\includegraphics{fig1.ps}

$$\vbox{\offinterlineskip
{
\def\tvi{\vrule height 12pt depth 5pt width 0pt}
\def\tv{\tvi\vrule}
\def\cc#1{\hfill\quad#1\quad\hfill}
\catcode`\*=\active \def*{\hphantom{0}}
\halign{\tv#&\cc{#}&\cc{#}&\cc{#}&\cc{#}&\tv#\cr
\noalign{\hrule}
&$ q $&$ C_{\rm o}^{\rm anal} $&$ C_{\rm o}^{\rm exp} $& Ref  &\cr
\noalign{\hrule}
& 20 & 5.362(3)& 5.2(2) & \ref{bil20}&\cr
\noalign{\hrule}
& 10 & 18.06(4)& 12.7(3) & \ref{bilcom}&\cr
&    &         & $\sim$ 18 & \ref{bilcom}&\cr
&    &         & 10.7(1.0) & \ref{koster}&\cr
\noalign{\hrule}
& *8 & 37.5(4)& 23.(3.) & \ref{koster}&\cr
\noalign{\hrule}
& *7 & 71.3(1.0)& 47.5(2.5) & \ref{janke}&\cr
&    &         & 50.(10.)  & \ref{bilcom}&\cr
&    &         & 44.4(2.2)  & \ref{rummu}&\cr
\noalign{\hrule}
}}}$$
\centerline{\bf Table 1}
\noindent{\it The ordered phase specific heat, our prediction compared
to numerical estimates}
\bigskip
At $q=20$ our prediction is in good agreement with the numerical estimate.
In contrast it strongly disagrees at $q\leq 10$ with the value of $C_{\rm o}$
obtained from a finite size analysis of the maximum of the specific heat
measured in the coexistence regime. However it agrees at $q=10$ with
the estimate obtained in \ref{bilcom} at $\beta=\beta_{\rm t}$, in accordance
with the rigorous statements of \ref{boko}.
\bigskip
The large $q$ expansion analysis supports the idea that not only does
the correlation length in a pure phase diverge at $\beta=\beta_{\rm t}$,
$q\rightarrow 4_+$, but also that the associated fluctuations imply
divergences of the energy cumulants. For example we obtain
$F_{\rm o}^{(3)} \sim -1800$ at $q=10$ (see \fig{1}).
Moreover the internal energy fluctuations behave
in a way consistent with $C / \xi$ being finite at $q=4_+$
as it is known to be at  $q=4_-$.
We then propose the following {\it ansatz}:
\item  {}There exists a region of $q>4$ where the free energies around
$\beta=\beta_{\rm t}$ reflect accurately the scaling properties associated with
the second-order point lying at $q=4 , \beta = \ln 3$ and characterized
 by the corresponding critical indices $\alpha$ and $\nu$.

\noindent Specifically, according to the known value 2/3 of $\alpha$,
we parametrize $F_{\rm o}^{(2)}(\beta)$  at
$q=4$ and for $\beta\rightarrow (\beta_{\rm t})_+$ as
$$F_{\rm o}^{(2)}(\beta)=A \ (\beta-\beta_{\rm t})^{-2/3} . \EQNO{F2S}$$
The higher derivatives are trivially deduced and in their
expressions we replace $(\beta-\beta_{\rm t})$ by ( Cst ~ $\xi^{-3/2}$ )
from \eq{xi4} and, boldly continuing above $q=4$ at $\beta=\beta_{\rm t}$,
reexpress $F^{(p+2)}_{\rm o}$ as a function of $x$ via \eq{xi} to get
$$F^{(p+2)}\equiv F_{\rm o}^{(p+2)}(\beta_{\rm t})=A \ (-)^p
 {\Gamma(2/3+p)\over\Gamma(2/3)} (Bx)^{1+3p/2} , \EQNO{Fx}$$
where the constant $B$ takes into account proportionality constants.
Thus we get the following representation for the ordered phase free energy
$$F_{\rm o}(\beta)=F(\beta_{\rm t})-E_{\rm o} (\beta-\beta_t) +
 \sum_{n=2}^{\infty} (\beta-\beta_{\rm t})^n {F^{(n)}\over n!} \EQNO{FoB}$$
where we introduced the known linear term ($F^{(1)}=-E_{\rm o}$), and took
$F^{(n)}$ as given by \eq{Fx} for $n>1$.
Note that all the odd cumulants are negative, as we found to be the case
for $F^{(3)}$ from the large $q$ expansion.
A similar expression holds for $F_{\rm d}$, starting from \eq{F2S} with
$(\beta-\beta_{\rm t}) \rightarrow (\beta_t-\beta)$, and replacing
 $E_{\rm o}$ by $E_{\rm d}$ in \eq{FoB}.

It is easy to sum the series \eq{FoB}. For later convenience, we introduce
scaled temperature ($v$), energy ($\epsilon$) and length (${\cal S}$) variables
$$v=(\beta-\beta_{\rm t}) (B x)^{3/2}\qquad ,\qquad
\epsilon={E (Bx)^{1/2}\over 3A} \qquad ,\qquad
{\cal S}^2={(Bx)^2\over 3A} \EQNO{scaled}$$
and end up with the following compact result
$$F_{\rm o}(\beta)=F(\beta_{\rm t})+{1\over {\cal S}^2} \bigl [
 -{3\over4}-(\epsilon_{\rm o}+1) v +{3\over4}(1+v)^{4/3} \bigr] \EQNO{final}$$
This equation is the central result of this letter. We claim that, although we
neglected all regular and less singular contributions to \eq{F2S}, \eq{final}
summarizes accurately, over a wide range of $q>4$ values, all the properties of
the model.

Let us justify this statement by comparing the predictions of \eq{final}
for the energy distribution to data \ref{bilcom} taken at $q=10$ with
various $L$ values. This distribution is obtained by inverse Laplace
 transform of the partition function
$$ P_V(E)=N_1\int_{\beta_0-i\infty}^{\beta_0+i\infty}{\rm d}\beta \Bigl [
 q\exp[V F_{\rm o}(\beta)]+\exp[V F_{\rm d}(\beta)]
\Bigr ] \exp[VE(\beta-\beta_{\rm t})] \EQNO{PV}$$
where $F_{\rm d}$ follows from \eq{final} by $v\rightarrow-v$ and
$\epsilon_{\rm o}\rightarrow\epsilon_{\rm d}$ (this is consistent with
duality up to terms of order $x^{-3/2}$ as compared to 1) and
with $N_1$ a factor ensuring
\vfill\eject
{}~~
\vskip 15.5 truecm
\includegraphics{fig2.ps}

\noindent the probability normalization.
Trading $(\beta-\beta_{\rm t})$ for $v$ and $E$ for $\epsilon$
of \eq{scaled}, we get
$$P_V^{\rm o}(E)=N_2\int_{v_{\rm o}-i\infty}^{v_{\rm o}+i\infty}
{\rm d}v\exp\bigl[({L\over{\cal S}})^2 [(\epsilon-\epsilon_{\rm o}-1)v
+{3\over4}(1+v)^{4/3})]\bigr] \EQNO{PVo}$$
Details on the computation of this integral will be given in \ref{futur}
and here we limit ourselves to short remarks.
\item{i)} As a function of $\epsilon-\epsilon_{\rm o}$, $P_V^{\rm o}$
depends on $q$ and $L$
only through the scaled volume $({L / {\cal S}})^2$.
\item{ii)} Any $v_0>-1$ is suitable and the integral converges exponentially.
\vfill\eject
{}~~
\vskip 15.5 truecm
\includegraphics{fig3.ps}

\item{iii)} At large $(L/{\cal S})$, a saddle point method can be valuable.
However $(L/{\cal S})$ is not large in practice and there exists an energy
value slightly above $E_{\rm o}$  where the saddle point value reaches $v=-1$
(metastability point).
\item{iv)} Actual computation requires numerical values of $A$ and $B$. At
$q=10$ we get $A=.193$ and $B=.386$ from the large $q$ expansion results
(at other $q$ values, slight changes have to be made according to \fig{1} ).
\item{v)} At $q=10$ the length scale is ${\cal S}\sim 60$,
nearly 6 times the correlation length, so that for current values of $L$
the ratio $(L/{\cal S})$ is hardly of order 1 !

For the above reasons, we compute the integral \eq{PVo} numerically.
The results are shown together with the data of \ref{bilcom} on \fig{2}
for $L=16,20,24$, and on \fig{3} for $L=36,44,50$.
Remembering that the continuous curves are absolute predictions
without any free parameter, it is quite striking to see how such a simple
ansatz as \eq{PVo} yields good results. Because $(L/{\cal S})$ is not large,
they are very different from what an asymptotic expansion would give.
For example $E_{\rm peak}-E_{\rm o}$ behaves effectively as $\sim 1/L$
over a wide range of $L$ values, whereas the asymptotic expectation (saddle
point method in \eq{PV}) is

$L^2(E_{\rm o}-E_{\rm peak})=-F_{\rm o}^{(3)} / ( 2F_{\rm o}^{(2)})\sim
 15,100,350,1000 \quad{\rm at}\quad q=20,10,8,7$.

Marked discrepancies only appear at the external edges of the ordered (left)
and disordered (right) peaks. In particular the theoretical curve levels out
unduly at the bottom right of \fig{2}. This is a spurious effect :
the tail of the ordered peak contributes more than the disordered phase.
However it appears at a negligibly small level at larger $L$'s,
being asymptotically of order $\exp[-c \ L^2]$.
As $L$ is increased (\fig{3}), the agreement between predictions and data
becomes better and better at nearly all values of $E$, but around the dip
between the peaks. There indeed mixed phase contributions with percolating
interfaces should finally win over the pure phase contributions.
We consider the small departure of the theoretical
curve below the data points at $L=50$ as an evidence for the emergence of
mixed phase contributions. Since the dip region is often
used for the determination of the interface tension $\sigma$ because
 strip configurations eventually yield an
energy independent plateau in $P_V(E)$ with \ref{binder}
$$2 \ \sigma \simeq -{1\over L} \ln {P_V(E_{{\rm dip}}) \over
 P_V(E_{{\rm peak}})} , \EQNO{2sig}$$
the value and $L$ dependence of the right hand side of \eq{2sig} are
interesting issues. Although in our construction this quantity diverges as $L$
asymptotically, we unexpectedly find it roughly constant for $L\leq 50$,
at a value around .11, not very far from the exact value
$1/\xi_{\rm d}=0.95$ \ref{bogs}.
This casts some doubt on attempts to determine $\sigma$ from \eq{2sig}
\refand{janke}{rummu} unless the plateau in $E$ is seen
\refand{plateau}.

Summarizing, we have shown, for the 2-d Potts model at $q>4$, up to large $q$
values, that the bulk properties are accurately described close to
$\beta_{\rm t}$ by free energies $F_{\rm o}(\beta)$
and $F_{\rm d}(\beta)$ inferred by a simple ansatz from the known properties
of the second-order point at $q=4$.
Definite expressions for $F_{\rm o}(\beta)$ and $F_{\rm d}(\beta)$
were obtained by matching this ansatz to a calculation of their large $q$
expansion, achieved at order $(1/\sqrt{q})^9$.
Our main result is expressed in \eq{final} and illustrated by the adequacy
of the corresponding energy distribution to explain most of the features
observed in numerical simulations at finite volume.

In the first-order region, each phase knows little about
the existence of the other ones, and rather feels the effectively close
critical point $q=4,\beta=\beta_{\rm t}$, from which it inherits nice
scaling properties. This is so as long as the linear size of the box
is smaller than the length scale ${\cal S}$, proportional to the correlation
length $\xi$, but many times larger ( ${\cal S}= 5 \ {\rm to} \ 7 \xi$ in
the range $q=7 \ {\rm to} \ 20$).
Only at $V\geq{\cal S}^2$ asymptotics takes place, so that interface tension
effects can show up and be measured.

Similar ideas could be applied to other situations. One such situation is
the 3-d Potts model at $q\geq3$ where the first-order regime could
also be influenced by the second-order point $q_{3 \rm d}$ situated
between $q=2$ (Ising) and $q=3$. Then one
might get some information on this critical point from numerical studies
at $q\geq3$. Another interesting case is QCD at finite temperature;
although the transition is first-order, universality might be nevertheless
invoked to relate its behaviour to that of the 3-d 3 states Potts model,
both models sharing the same (universal) behaviour at
a "close" point of parameter space.

We are aware of the fact that our ansatz describes the pure phase free energies
as analytic functions of $\beta$ at $\beta_{\rm t}$, which they probably are
not, in the same way as in field driven first-order transitions the zero-field
point is an essential singularity of the free energy \ref{isak}.
This question deserves further discussion, but we believe that our ansatz,
although not `analytically correct', takes into account the most significant
 features of the model.

It is a pleasure to thank R. Balian and A. Billoire for illuminating
 discussions.
We also acknowledge useful conversations with B. Grossmann and T. Neuhaus.
\vfill\eject
\bigskip\centerline{\sectnfont References}\bigskip
\item {\reftag{boko})} C.~Borgs and R.~Koteck\'y, J. Stat. Phys. 61 (1990) 79;
C.~Borgs, R.~Koteck\'y and S.~Miracle-Sol\'e, J. Stat. Phys. 62 (1991) 529.
\item{\reftag{wu})} F.~Y.~Wu, Rev. Mod. Phys. 54 (1982) 235.
\item{\reftag{baxt})} R.~J.~Baxter, J. Phys. C6 (1973) L445.
\item{\reftag{xi})} A.~Kl\"umper, A.~Schadschneider and J.~Zittartz,
Z Phys. B76 (1989) 247.
\item \    E.~Buffenoir and S.~Wallon, to appear in Journal of Phys. A.
\item{\reftag{bogs})} C.~Borgs and W.~Janke,
{\it An explicit formula for the interface tension of the 2D Potts Model}
Preprint FUB-HEP 13/92, HRLZ 54/92.
\item{\reftag{kast})} P.~W.~Kasteleyn and C.~M.~Fortuin, J. Phys. Soc. Japan
{\bf 26} (Suppl.), 11 (1969).
\item{\reftag{futur})} T.~ Bhattacharya, R.~Lacaze and A.~Morel,
 in preparation.
\item{\reftag{enting})} I.~G.~Enting, J. Phys. A10 (1977) 325.
\item {\reftag{challa})} M.~S.~S.~Challa, D.~P.~Landau and K.~Binder,
      Phys. Rev. B34 (1986) 1841.
\item {\reftag{bil20})} A.~Billoire, T.~Neuhaus and B.~Berg,
Nucl. Phys. B(in press).
\item {\reftag{bilcom})} A.~Billoire, R.~Lacaze and A.~Morel,
Nucl. Phys. B 370 (1992) 773.
\item{\reftag{koster})} J.~Lee and J.~M.~Kosterlitz,
 Phys. Rev. B43 (1991) 3265.
\item {\reftag{janke})}  W.~Janke, B.~Berg and M.~Katoot,
 Nucl. Phys. B382 (1992) 649.
\item {\reftag{rummu})}  K.~Rummukainen,
{\it Multicanonical Cluster Algorithm and the 2-D 7-State Potts Model},
Preprint CERN-TH.6654/92.
\item {\reftag{binder})} K.~Binder, Phys. Rev A25 (1982) 1699;
Z. Phys. B43 (1981) 119.
\item {\reftag{isak})} M.~E.~Fisher, Physics (N.Y.) 3 (1967) 255;
S.~N.~Isakov, Comm. Math. Phys. 95 (1984) 427.
\item {\reftag{plateau})} B.~Berg, U.~Hansmann and T.~Neuhaus,
to appear in Proceedings
"Computer Simulations Studies in Condensed Matter Physics", Athens 1992;
\item {} B.~Grossmann and M.~L.~Laursen, {\it The Confined-Deconfined Interface
Tension in Quenched QCD using the Histogram Method}, Preprint HLRZ-93-7;
\item {} A.~Billoire, T.~Neuhaus and B.~Berg, {\it A Determination of Interface
Free Energies}, in preparation.
\vfil\end